# Reliance on Metrics is a Fundamental Challenge for AI

**Rachel L. Thomas**

University of San Francisco
rlthomas3@usfca.edu

**David Uminsky**

University of San Francisco
duminsky@usfca.edu

**Abstract**

Optimizing a given metric is a central aspect of most current AI approaches, yet overemphasizing metrics leads to manipulation, gaming, a myopic focus on short-term goals, and other unexpected negative consequences. This poses a fundamental challenge in the use and development of AI. We first review how metrics can go wrong in practice and aspects of how our online environment and current business practices are exacerbating these failures. We put forward here an evidence based framework that takes steps toward mitigating the harms caused by overemphasis of metrics within AI by: (1) using a slate of metrics to get a fuller and more nuanced picture, (2) combining metrics with qualitative accounts, and (3) involving a range of stakeholders, including those who will be most impacted.

## Introduction

Metrics can play a central role in decision making across data driven organizations and their advantages and disadvantages have been widely studied (Likierman 2009, Kaplan and Norton 1992). Metrics play an even more central role in AI algorithms and as such their risks and disadvantages are heightened. Some of the most alarming instances of AI algorithms run amok, such as recommendation algorithms contributing to radicalization (Ribeiro et. al. 2019), teachers described as "creative and motivating" being fired by an algorithm (Tuque 2012), or essay grading software that rewards sophisticated garbage (Ramineni and Williamson 2018) all result from over-emphasizing metrics. We have to understand this dynamic in order to understand the urgent risks we are facing due to misuse of AI.

At their heart, what most current AI approaches do is to optimize metrics. The practice of optimizing metrics is not new nor unique to AI, yet AI can be particularly efficient (even too efficient!) at doing so. This unreasonable effectiveness at optimizing metrics results in one of the grand challenges in AI design and ethics: the metric optimization central to AI often leads to manipulation, gaming, a focus on short-term quantities (at the expense of longer-term concerns), and other undesirable consequences, particularly when done in an environment designed to exploit people's impulses and weaknesses. Moreover, this challenge also yields, in parallel, an equally grand contradiction in AI development: optimizing metrics results in far from optimal outcomes.

Some of the issues with metrics are captured by Goodhart's Law, "When a measure becomes a target, it ceases to be a good measure" (Goodhart 2015 and Strathern 1997). We examine in this paper, through a review of a series of real-world case studies, how Goodhart's Law as well as additional consequences of AI's reliance on optimizing metrics is already having an impact in society. We find from this review the following well supported principles:

- Any metric is just a proxy for what you really care about
- Metrics can, and will, be gamed
- Metrics tend to overemphasize short-term concerns
- Many online metrics are gathered in highly addictive environments

Given the harms of over-emphasizing metrics, it is important to work to mitigate these issues which will remain in most use cases of AI. We conclude by proposing a framework for the healthier use of metrics that includes:

- Use a slate of metrics to get a fuller picture
- Combine metrics with qualitative accounts
- Involve a range of stakeholders, including those who will be most impacted

# Goodhart's Law and Machine Learning

Goodhart's Law is often given as one of the limitations of metrics: a measure (or metric) that becomes a target ceases to be a good measure. This is a fundamental challenge with metrics. The formal mathematical definition of a metric (Rosenlicht, 1968) is a rule that, for each pair of elements *p* and *q* in a set *E*, associates to them a real-number *d(p,q)*, which has the following properties:

- *d(p,q) ≥ 0*
- *d(p,q) = 0 if and only if p = q*
- *d(p,q) = d(q,p) for all p* and *q* in the set
- *d(p,r) ≥ d(p,q) + d(q,r)*

A metric can be used to measure the magnitude of any element in the set for which it is defined, by evaluating the distance between that element and zero.

This concept of a metric for an organization is often used interchangeably with key performance indicators (KPI) which can be defined as "the quantifiable measures an organization uses to determine how well it meets its declared operational and strategic goals" (Schrage 2018). In practice metrics and KPIs are also used more generally to refer to measurements made in the realm of software products or business, such as page views/impressions, click through rates, time spent watching, quarterly earnings, and more. Here, metric is used to refer to something which can be measured and quantified as a numeric value. Although they are not identical, the mathematical definition of metric (which is relevant for formal machine learning work that involves optimization for a cost or loss function) is very much linked to the more informal business usage of the term metric.

Machine learning has been defined as follows: "A computer program is said to learn from experience *E* with respect to some class of tasks *T* and performance measure *P*, if its performance at tasks in *T*, as measured by *P*, improves with experience *E*" (Mitchell 1997, as quoted in Goodfellow, Bengio, and Courville 2016). Here, the performance measure *P* must be a well-defined quantitative measure, such as accuracy or error rate. Defining and choosing this measure can involve a number of choices, such as whether to give partial credit for some answers, how to weigh false positives relative to false negatives, the penalties for frequent medium mistakes relative to rare large mistakes, and more.

In the context of deep learning, according to Goodfellow, Bengio, and Courville (2016), training a model is the process of:

> finding the parameters $\theta$ of a neural network that significantly reduce a cost function $J(\theta)$, which typically includes a performance measure evaluated on the entire training set as well as additional regularization terms

That is, model training is explicitly defined around this process of optimizing a particular metric (in this case, minimizing cost).

Goodhart's Law, that a measure that becomes a target ceases to be a good measure, remains a relevant lens to see the shortcomings of AI's reliance on metrics. This law has appeared in various forms since economist Charles Goodhart first proposed it in 1975 in response to how monetary metrics broke down after central banks adopted them as targets. This arose out of attempts in the early 1970s to control inflation by choosing metrics with a stable relationship to inflation as targets. Central banks from a number of countries did this, and in most cases the relationship between inflation and the metric chosen broke down once that metric became a target. In his entry on Goodhart's Law in The Encyclopedia of Central Banking (Goodhart 2015), Goodhart refers to the popular phrasing of his namesake law, "When a measure becomes a target, it ceases to be a good measure" (Strathern 1997). Note that the target of training a deep learning model is often defined as improving or optimizing our cost function $J(\theta)$. That is, by definition, *deep learning is a process in which a measure is the target*. Thus, there is an increased likelihood that any risks of optimizing metrics are heightened by AI (Slee 2019), and Goodhart's Law grows increasingly relevant.

# Previous related work

As part of a much broader survey of the field of AI ethics, Green (2018), refers to concerns of "Goodharting" the explanation function; for instance, if an algorithm learns to produce answers that humans find appealing as opposed to accurate answers.

Work by Manheim and Garrabrant (2019) develops a taxonomy of 4 distinct failure modes that are all contained in Goodhart's Law. Here Goodhart's law is framed in terms of a metric being chosen as a proxy for a goal, and the collapse that occurs with that proxy as a target: (1) regressional Goodhart, in which the difference between the proxy and the goal is important, (2) extremal Goodhart, in which the relationship between the proxy and the goal is different when taken to an extreme, (3) causal Goodhart, in which there is a non-causal relationship between the proxy and goal, and intervening on one may fail to impact the other, (4) adversarial Goodhart, in which adversaries attempt to game the chosen proxy. The authors then provide subcategories within causal Goodhart, based off of different underlying relationships between the proxy and the goal, and different subcategories of adversarial Goodhart, based on the behavior of the regulator and the agent. Mathematical formulations for the failure modes are presented.

A related issue is "reward hacking" in which an algorithm games its reward function, learning a "clever"

way of optimizing the function which subverts the designer/programmer's intent. This is most often talked about in the field of reinforcement learning. For example, in a reimplementation of AlphaGo, the computer player learned to pass forever if passing was an allowed move. Victoria Krakovna maintains a spreadsheet with dozens of such examples (Krakovna 2019). While this is a subset of the type of cases we will consider, reward hacking typically refers to the algorithm's internal behavior, whereas here we are primarily interested in systems that involve a mix of human and algorithmic behavior.

The issue of over optimizing single metrics is so worrisome that recent work (Taylor 2016) has suggested replacing direct optimization with "q-quantilizers" which practically attempt to replace offering the deterministic highest utility maximizer with random alternatives weighted by their utility. The idea purports to mitigates Goodharting but at the cost of lower utility and is sensitive to how well understood what a "safe" base distribution of actions looks like. Slee (2019) goes further to state that in many cases AI's reliance on pure incentive maximization is incompatible with actually achieving optimal outcomes for the intended use cases.

Our goal here is to elucidate different characteristics of how Goodhart's Law manifests in a series of real-world cases studies, aspects of how our online environment and current business practices are exacerbating these failures, and propose a framework towards mitigating the harms caused by overemphasis of metrics within AI.

## We can't measure the things that matter most

Metrics are typically just a proxy for what we really care about. Mullainathan and Obermeyer (2017), cover an interesting example: the researchers investigate which factors in someone's electronic medical record are most predictive of a future stroke and find that several of the most predictive factors (such as accidental injury, a benign breast lump, or colonoscopy) don't make sense as risk factors for stroke. It turned out that the model was simply identifying people who utilize health care a lot. They didn't actually have data of who had a stroke (a physiological event in which regions of the brain are denied new oxygen); they had data about who had access to medical care, chose to go to a doctor, were given the needed tests, and had this billing code added to their chart. But a number of factors influence this process: who has health insurance or can afford their co-pay, who can take time off of work or find childcare, gender and racial biases that impact who gets accurate diagnoses, cultural factors, and more. As a result, the model was largely picking out people who utilized healthcare versus who did not.

This an example of the common phenomenon of having to use proxies: You want to know what content users like, so you measure what they click on. You want to know which teachers are most effective, so you measure their students test scores. You want to know about crime, so you measure arrests. These things are not the same. Many things we do care about cannot be measured. Metrics can be helpful, but we can't forget that they are just proxies.

As another example, Google used hours spent watching YouTube as a proxy for how happy users were with the content, writing on the Google blog that "If viewers are watching more YouTube, it signals to us that they're happier with the content they've found" (Meyerson 2012). Guillaume Chaslot, founder of independent watch group AlgoTransparency and an AI engineer who formerly worked at Google/YouTube, shares how this had the side effect of incentivizing conspiracy theories, since convincing users that the rest of the media is lying kept them watching more YouTube (Chaslot 2018).

## Metrics can, and will, be gamed

It is almost inevitable that metrics will be gamed, particularly when they are given too much power. A detailed case study about "a system of governance of public services that combined targets with an element of terror" developed for England's health care system in the 2000s covers many ways that gaming can occur (Bevan and Hood 2006). The authors identify three well-documented forms of gaming from Soviet era production targets: (1) ratchet effects: 'a wise director fulfils the plan 105%, but never 125%', (2) threshold effects: crowd distribution towards target, (3) output distortions: achieve targets at the cost of significant but unmeasured aspects of performance. They then detail many specific examples of how gaming manifested in the English healthcare system. For example, targets around emergency department wait times led some hospitals to cancel scheduled operations in order to draft extra staff to the emergency room, to require patients to wait in queues of ambulances, and to turn stretchers into "beds" by putting them in hallways. There were also significant discrepancies in numbers reported by hospitals versus those reported by patients; for instance, around wait times, according to official numbers 90% of patients were seen in less than 4 hours, but only 69% of patients said they were seen in less than 4 hours when surveyed (Bevan and Hood 2006).

As education policy in the United States began over-emphasizing student test scores as the primary way to evaluate teachers, there have been widespread scandals of teachers and principals cheating by altering students' scores, in Georgia, Indiana, Massachusetts, Nevada, Virginia, Texas, and elsewhere (Gabriel 2010). One consequence of this is that teachers who don't cheat may be penalized or even fired, when it appears student test scores have dropped to more average levels under their instruction (Turque 2012). When metrics are given undue importance, attempts to game those metrics become common.

A modern AI case study can be drawn from recommendation systems, which are widely used across many platforms to rank and promote content for users. Platforms are rife with attempts to game their algorithms, to show up higher in search results or recommended content, through fake clicks, fake reviews, fake followers, and more (Tufekci 2019). There are entire marketplaces for purchasing fake reviews and fake followers, etc.

During one week in April 2019, Chaslot collected 84,695 videos from YouTube and analyzed the number of views and the number of channels from which they were recommended (Harwell and Timberg 2019).
The state-owned media outlet Russia Today, abbreviated RT, was an extreme outlier in how much YouTube's algorithm had selected it to be recommended by a wide-variety of other YouTube channels. According to Harwell (2019):

> Chaslot said in an interview that while the RT video ultimately did not get massive viewership — only about 55,000 views — the numbers of recommendations suggest that Russians have grown adept at manipulating YouTube's algorithm, which uses machine-learning software to surface videos it expects viewers will want to see. The result, Chaslot said, could be a gradual, subtle elevation of Russian views online because such videos result in more recommendations and, ultimately, more views that can generate more advertising revenue and reach.

Automatic essay grading software currently used in at least 22 USA states focuses primarily on metrics like sentence length, vocabulary, spelling, and subject-verb agreement, but is unable to evaluate aspects of writing that are hard to quantify, such as creativity. As a result, gibberish essays randomly generated by computer programs to contain lots of sophisticated words score well. Essays from students in mainland China, which do well on essay length and sophisticated word choice, received higher scores from the algorithms than from expert human graders, suggesting that these students may be using chunks of pre-memorized text (Ramineni and Williamson, 2018).

## Metrics overemphasize short-term concerns

It is much easier to measure short-term quantities: click through rates, month-over-month churn, quarterly earnings. Many long-term trends have a complex mix of factors and are tougher to quantify. While short-term incentives have led YouTube's algorithm to promote pedophilia (Fisher and Taub 2019), white supremacy (Ribeiro et. al. 2019), and flat-earth theories (Landrum 2018), the long-term impact on user trust will not be positive. Similarly, Facebook has been the subject of years worth of privacy scandals, political manipulation, and facilitating genocide (Vaidhyanathan 2018), which is now having a longer-term negative impact on Facebook's ability to recruit new engineers (Bowles 2018).

Simply measuring what users click on is a short-term concern, and does not take into account factors like the potential long-term impact of a long-form investigative article which may have taken months to research and which could help shape a reader's understanding of a complex issue and even lead to significant societal changes.

The Wells Fargo account fraud scandal provides a case study of how letting metrics replace strategy can harm a business (Harris and Tayler 2019). After identifying cross-selling as a measure of long-term customer relationships, Wells Fargo went overboard emphasizing the cross-selling metric: intense pressure on employees combined with an unethical sales culture led to 3.5 million fraudulent deposit and credit card accounts being opened without customers' consent. The metric of cross-selling is a much more short-term concern compared to the loftier goal of nurturing long-term customer relationships. Overemphasizing metrics removes our focus from long-term concerns such as our values, trust and reputation, and our impact on society and the environment, and myopically focuses on the short-term.

## Many metrics gather data of what we do in highly addictive environments

It matters which metrics we gather and in what environment we do so. Metrics such as what users click on, how much time they spend on sites, and "engagement" are heavily relied on by tech companies as proxies for user preference, and are used to drive important business decisions. Unfortunately, these metrics are gathered in environments engineered to be highly addictive, laden with dark patterns, and where financial and design decisions have already greatly circumscribed the range of options.

While this is not a characteristic inherent to metrics, it is a current reality of many of the metrics used by tech companies today. A large-scale study analyzing approximately 11K shopping websites found 1,818 dark patterns present on 1,254 websites, 11.1% of the total sites (Mathur, et. al. 2019). These dark patterns included obstruction, misdirection, and misrepresenting user actions. The study found that more popular websites were more likely to feature these dark patterns.

Zeynep Tufekci compares recommendation algorithms (such as YouTube choosing which videos to auto-play for you and Facebook deciding what to put at the top of your newsfeed) to a cafeteria shoving junk food into children's faces (Lewis 2018):

> This is a bit like an autopilot cafeteria in a school that has figured out children have sweet-teeth, and also like fatty and salty foods. So you make a line offering such food, automatically loading the next plate as soon as the bag of chips or candy in front of the young person has been consumed.

As those selections get normalized, the output becomes ever more extreme: "So the food gets higher and higher in sugar, fat and salt – natural human cravings – while the videos recommended and auto-played by YouTube get more and more bizarre or hateful."

Too many of our online environments are like this, with metrics capturing that we love sugar, fat, and salt, not taking into account that we are in the digital equivalent of a food desert and that companies haven't been required to put nutrition labels on what they are offering. Such metrics are not indicative of what we would prefer in a healthier or more empowering environment.

## A Framework for a Healthier Use of Metrics

All this is not to say that we should throw metrics out altogether. Data can be valuable in helping us understand the world, test hypotheses, and move beyond gut instincts or hunches. Metrics can be useful when they are in their proper context and place. We propose a few mechanisms for addressing these issues:

- Use a slate of metrics to get a fuller picture
- Combine with qualitative accounts
- Involve a range of stakeholders, including those who will be most impacted

**Use a slate of metrics to get a fuller picture and reduce gaming**

One way to keep metrics in their place is to consider a slate of many metrics for a fuller picture (and resist the temptation to try to boil these down to a single score). For instance, knowing the rates at which tech companies hire people from under-indexed groups is a very limited data point. For evaluating diversity and inclusion at tech companies, we need to know comparative promotion rates, cap table ownership, retention rates (many tech companies are revolving doors driving people from under-indexed groups away with their toxic cultures), number of harassment victims silenced by NDAs, rates of under-leveling, and more. Even then, all this data should still be combined with listening to first-person experiences of those working at these companies.

Likierman wrote in Harvard Business Review (2009) that using a diverse slate of metrics is one strategy to avoid gaming:

> It helps to diversify your metrics, because it's a lot harder to game several of them at once. [International law firm] Clifford Chance replaced its single metric of billable hours with seven criteria on which to base bonuses: respect and mentoring, quality of work, excellence in client service, integrity, contribution to the community, commitment to diversity, and contribution to the firm as an institution.

Likierman (2009) also cites the example of Japanese telecommunications company SoftBank using performance metrics defined for three distinct time horizons to make them harder to game.

**Combine with Qualitative Accounts**
Columbia professor and New York Times Chief Data Scientist Chris Wiggins wrote that quantitative measures should always be combined with qualitative information, "Since we cannot know in advance every phenomenon users will experience, we cannot know in advance what metrics will quantify these phenomena. To that end, data scientists and machine learning engineers must partner with or learn the skills of user experience research, giving users a voice" (Wiggins 2018).

Proposals such as Model Cards for Model Reporting (Mitchell, et. al. 2019) and Datasheets for Datasets (Gebru 2018) can be viewed in line with this thinking. These works acknowledge that the metrics typically accompanying models, such as performance on a particular dataset, are insufficient to cover the many complex interactions that can occur in real-world use, as the model is applied to different populations and in different use cases, and as the use potentially veers away from the initial intent. Mitchell (2019) and Gebru (2018) propose documenting much richer and more comprehensive details about a given model or dataset, including more qualitative aspects, such as intended use cases, ethical considerations, underlying assumptions, caveats, and more.

**Involve a range of stakeholders, including those who will be most impacted**

Another key to keeping metrics in their proper place is to keep domain experts and those who will be most impacted closely involved in their development and use. Empowering a diverse group of stakeholders to understand the implications and underlying assumptions of AI models is one of the goals of Model Cards for Model Reporting (Mitchell et al. 2018). We suggest going even further and including these stakeholders in the initial development process of these metrics in the first place.

Tool 3 in the Markkula Center's Ethical Toolkit for Engineering/Design Practice (Vallor, Green, and Raicu 2018) is to "expand the ethical circle" to include the input of all stakeholders. They suggest a number of questions to ask about this topic, including:

- Whose interests, desires, skills, experiences and values have we simply assumed, rather than actually consulted? Why have we done this, and with what justification?
- Who are all the stakeholders who will be directly affected by our product? How have their interests been protected? How do we know what their interests really are—have we asked?
- Who/which groups and individuals will be indirectly affected in significant ways? How have their interests been protected? How do we know what their interests really are—have we asked?

While their focus is on tech policy, the Diverse Voices paper (Young, Magassa, and Friedman 2019) provides a detailed methodology on how to elicit the expertise and feedback of underrepresented populations that would be useful in improving the design and implementation of metrics.

An additional argument for the importance of including a range of stakeholders comes from the fragile and elastic nature of deep learning algorithms combined with the incompatible incentives the owners of these algorithms have to address this fragility (Slee 2019). Specifically:

> …if subjects follow their incentives then the algorithm ceases to function as designed. To sustain their accuracy, algorithms need external rules to limit permissible responses. These rules form a set of guardrails which implement value judgments, keeping algorithms functioning by constraining the actions of subjects.

Slee proposes that external guardrails on how users are allowed to engage with the algorithms are a necessary corrective measure for the associated gaming and abuses cited above, but that the algorithm owners themselves are incentivized to create guardrails that don't align with their algorithms, an act of regulatory arbitrage. Given that guardrails are *a restriction on the user*, transparent and just corrective measures will depend on how effective the ethical circle has been expanded in the design and creation of the guardrails. Slee gives the example of the incompatible incentives Facebook faces in addressing ethical issues with its News Feed Algorithm, and suggests that journalists not tempted by the financial incentives driving Facebook would be better equipped to address this.

While it is impossible to simply oppose metrics, the harms caused when metrics are overemphasized include manipulation, gaming, a focus on short-term outcomes to the detriment of longer-term values, and other harmful consequences, particularly when done in an environment designed to exploit people's impulses and weaknesses, such as most of our online ecosystem. The unreasonable effectiveness of metric optimization in current AI approaches is a fundamental challenge to the field, and yields an inherent contradiction: solely optimizing metrics leads to far from optimal outcomes. However, we provide evidence in this paper that healthier use of metrics can be created by: (1) using a slate of metrics to get a fuller and more nuanced picture, (2) combining metrics with qualitative accounts, and (3) involving a range of stakeholders, including those who will be most impacted. This framework may help address the core paradox of metric optimization within AI, and not solely relying on metric optimization may lead to a more optimal use of AI.